\documentclass{article}
\usepackage{graphicx}
\title{Multifractal analysis and instability index of prior-to-crash market situations}

\begin{document}
\maketitle
\centerline{{\bf M.~Piacquadio} }
\centerline{Secretar\'ia de Investigaci\'on y Doctorado, Facultad de Ingenier\'ia, Universidad de Buenos Aires}
\centerline{Paseo Col\'on 850, Ciudad Aut\'onoma de Buenos Aires, Argentina} 
\vspace{12pt}
\centerline{ {\bf F. O. ~Redelico} }
\centerline{Laboratorio de Sistemas Complejos, Facultad de Ingenier\'ia, Universidad de Buenos Aires}
\centerline{Paseo Col\'on 850, Ciudad Aut\'onoma de Buenos Aires, Argentina} 
\centerline{Facultad de Ciencias Fisicomatem\'aticas e Ingenier\'ia, Universidad Cat\'olica Argentina}
\centerline{Alicia Moreau de Justo 1500, Ciudad Aut\'onoma de Buenos Aires, Argentina}

\begin{abstract}
We take prior-to-crash market prices (NASDAQ, Dow Jones Industrial Average) as a signal, a function of time, we project these discrete values onto a vertical axis, thus obtaining a Cantordust. We study said cantordust with the tools of multifractal analysis, obtaining spectra by definition and by lagrangian coordinates. These spectra have properties that typify the prior-to-crash market situation. Any of these spectra entail elaborate processing of the raw signal data. With the unprocessed raw data we obtain an instability index, also with properties that typify the prior-to-crisis market situation. Both spectra and the instability index agree in characterizing such crashes, and in giving an early warning of them.
\end{abstract} 

\section{Introducion}
We propose to apply some tools of multifractal analysis to the study of market crash situations (NASDAQ, Dow Jones Industrial Average). We take the data of market price fluctuations on a daily basis and we process it in two ways:
1) The prior-to-crash market prices are taken as a discrete one dimensional signal \textemdash this will be the Cantordust, covered with adjacent boxes $l_i$ with equal length $l$. These boxes are intervals. The weight $w_i$ of box $l_i$ will be the number of days in which the market prices are in box $l_i$. The weights $w_i$ are then normalized into probabilities $p_{r_i}$, by dividing each $w_i$ by the total $T$ cardinality of the sample. Then, the concentration $\alpha$ of a box $l_i$ is $\alpha(l_i) := \frac{\log p_{r_i}}{\log l}$. Therefore, $\alpha_{ min}$ corresponds to the heaviest box(es), $\alpha_{max}$ to the lightest one(s). The segment $[\alpha_{min}, \alpha_{max}]$ is divided into adjacent intervals of equal length $\Delta\alpha$, all $\alpha$'s in a $\Delta\alpha$ are identified, and $N_\alpha$ is their number. We define $f^*(\alpha) := \frac{\log N_\alpha}{\log 1/l}$. This spectrum-by-definition is hereby called $f^*(\alpha)$ in order to distinguish it from the usual lagrangian $f(\alpha)$ given by the thermodynamic algorithm: $\alpha = \tau '(q), f(\alpha) = q\alpha - \tau(q), q=f'(\alpha),f''(\alpha) < 0$, and $\tau(q) = \frac{\log ( \sum p_{r_i}^q)}{\log l}$, $l$ small, $q$ a parameter, $-\infty < q < \infty$.
\newline\indent
Notice that, in order to obtain $f^*(\alpha)$ and/or $f(\alpha)$, we \emph{process} weights $w_i$, $T$, and even $N_\alpha$, which is \textemdash for instance\textemdash \space always logarithmized and then divided by $log 1/l$ in order to produce $f^*(\alpha)$.
2) In this paper, we \textemdash also\textemdash \space use the \emph{unprocessed} data above: the \emph{unprocessed} $w_i$ and the unlogarithmized $N_\alpha$, in order to define a market instability  index as a quotient $Q_i$ that characterizes the prior-to-crash market situation. Then we use the \emph{processeed} $f^*(\alpha)$ and $f(\alpha)$ in order to strengthen the key properties of $Q_i$ that reveal the market collapse situation.
\begin{figure}
\begin{center}
\includegraphics[width=100mm]{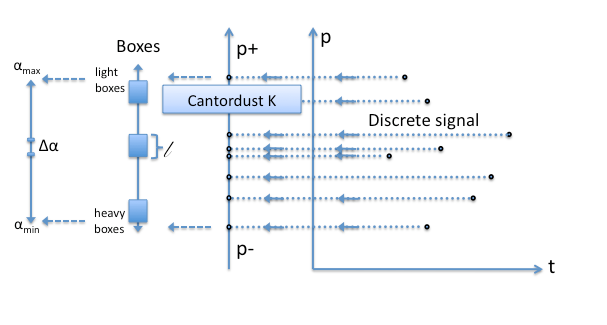}
\caption{Diagram D}
\end{center}
\end{figure}
\newline
\underline{Note}: $f^*(\alpha)$ and $f(\alpha)$ do not always coincide.
\section{A theoretical framework: the binomial case.}
We consider the simplest of Cantordusts: the ternary $K$ of Cantor in the unit interval $I = [0,1]$. We endow the subfractals \textemdash or "subcantors"\textemdash \space in $[0,\frac{1}{3}]$ and $[\frac{1}{3},1]$ with probability measures $p_r$ and $1-p_r$ respectively, $0 < p_r < 1$, $p_r \neq \frac{1}{2}$; $p_r^2$, $p_r(1-p_r)$, $(1-p_r)p_r$, and $(1-p_r)^2$ will be the weights of the $2^2$ subfractals of width $\frac{1}{3^2}$ in the second step of the construction of $K$, \dots and so on. This simplest of cases ascribes probability measure $p_r^{k-r}(1-p_r)^r$ to all subfractals of width $l = \frac{1}{3^k}$ in the $k$th step, $r$ between 0 and $k$.
\newline\indent
Then $\alpha$ (of such subfractal) $ = \frac{\log (prob.)}{\log (length)} = \frac{(k-r)\log p_r + r\log(1-p_r)}{\log{1/3^k}}= r\frac{\log{\frac{1-p_r}{p_r}}}{\log(1/3^k)} + \frac{\log p_r}{\log (1/3)} = Ar+B$, where $A=A_k$, and $r$ an integer, $r:0\rightarrow k$.
\newline\indent
The corresponding $f(\alpha)$, the same as $f^*(\alpha)$ in this simplest case, is $\frac{\log N\alpha}{\log 3^k}=\frac{\log {k\choose r}}{\log 3^k}$, where ${k \choose r}$ reaches its maximum for $r = \frac{k}{2}$ and is entirely symmetric for $r$ at left and right of $k/2$. The concentration $\alpha$ goes linearly with integer $r$ from an "origin" $B$ (where $f(\alpha)=0$) to $B+A \frac{k}{2}$, where $f(\alpha)$ reaches its maximum, to $B+A$, where $f(\alpha)$ is again zero. In this simplest binomial case, the increasing and decreasing branches of the curve are a mirror image of each other, like an inverted parabola. (In the case we work with below, such symmetry might be absent, still we will consider 
$k$ as a constant, $\alpha$ going linearly with integer $r:0 \rightarrow k$).
\newline\indent
We will simplify this binomial case further, in order to be able to adapt it to our study of the market crash signal. The binomial case comes from the unfolding of $(a+b)^k$, with $a=p_r$ and $b=1-p_r$, and the symmetric properties of ${k \choose r}$. We have a decreasing index for powers of $a$'s \textemdash i.e. for $p_r$'s\textemdash \space to wit: $p_r^{k}$, $p_r^{k-1}$, \dots coupled with an exponent increase for powers of $b$ \textemdash or $1 - p_r$\textemdash : $(1-p_r)^0,(1-p_r)^1$, \dots ; ${k \choose 0},{k \choose 1},$\dots the corresponding number of them. 
\newline\indent
Notice that the $\alpha$ \& $f(\alpha)$ structure above does not need the ternary of Cantor as a starting point: if we divide the unit invertal $I$ in halfs, in quarters, \dots where $l = \frac{1}{2^k}$ is the size of each interval-box in $k$th step, then we simply replace "3" by "2" in the denominators of $\alpha$ and $f(\alpha)$ above. In fact, if $L$ is the number of equal intervals dividing $I$, we still have $\alpha = Ar +B$, with $l=\frac{1}{L}$ in the denominator of $A$, $L$ in that of $B$, and $f(\alpha) = \frac{\log {k \choose r}}{\log L}$ just as symmetric as above. 
\newline\indent
Next, consider a number $P>1$ instead of $p_r$ above, and the binomial expression $(a+b)^k$, $a=P$ and $b=1$. We now have a decreasing sequence of weights $w_i$: $P^{k},P^{k-1},\dots,P^{k-r}$ \dots instead of $p_r^k,p_r^{k-1},\dots; 1-p_r$ has been replaced by 1. The total $T$ weight is now $T=(P+1)^k$. 
\newline\indent
Now, in order to make probabilities out of these weights $w_i$ we have to write $\frac{P^k}{(P+1)^k},\dots,\frac{P^{k-r}}{(P+1)^k},\dots$;$ {k \choose r}$ the corresponding number of them. With $L$ as before, $k$ large and \emph{fixed}, $T=(P+1)^k$ the size of the sample, or total number of such elements, we still have, as above, $\alpha= Ar+B$, $r:0\rightarrow k $: $\alpha = \frac{(k-r)\log P - \log T}{\log \frac{1}{L}}=\frac{\log(P+1)^k-k\log P+ r\log P}{\log L} = r\frac{\log P}{\log L} + k \frac{\log\frac{P+1}{P}}{\log L}$; and $f(\alpha) = \frac{\log {k \choose r}}{\log L}$ is just as symmetric as above. This will be our working structural framework.
\begin{figure}
\begin{center}
\includegraphics[width=90mm]{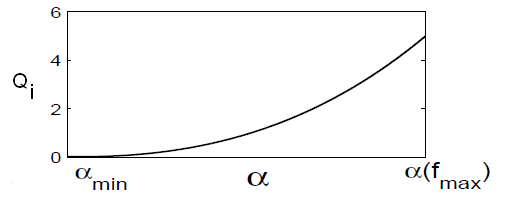}
\caption{Instability quotient, binomial case}
\end{center}
\end{figure}
\subsection{The instability quotient} 
$Q_i$ is a function of $\alpha$ or of $r$ indistinctly, and will be, by definition, $Q_i(\alpha) = Q_i(r):= \frac{N_\alpha}{P^{k-r}}=\frac{{k \choose r}}{P^{k-r}}$, an obviously increasing function of $r$, i.e. of $\alpha$, between $\alpha_{min}$ \textemdash or $r=0$\textemdash \space and $\alpha(f_{max})$ \textemdash or $r=\frac{k}{2}$ in our binomial case.
\newline\indent
Notice that the instability quotient is made up of "unprocessed" weights \textemdash not logarithms of probabilities\textemdash \space and "unlogarithmized" $N_\alpha$, i.e. with the raw data which, once processed, yields $f^*(\alpha)$ and $f(\alpha).$
\begin{figure}
\begin{center}
\includegraphics[width=90mm]{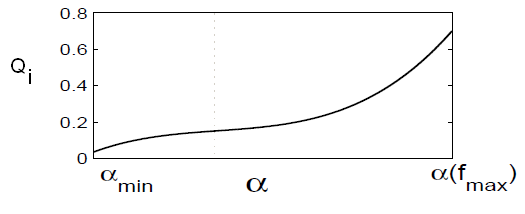}
\caption{Instability quotient, NQ1000}
\end{center}
\end{figure}
\section{Structural frame for the study of market crash  situations}
NQ1000 refers to the 1000 daily fluctuations before the April 2000 NASDAQ crash; DJIA1600, DJIA900, \dots etc. refer to the 1600, 900, \dots etc. daily fluctuations before the October 1997 Dow Jones market crash, i.e. 1600 days prior to crash, 900 days ditto, etc. (Rotundo, 2006). We start with NQ1000, and we represent (ditto for other cases)  the graph of the closing market price $p$ as a function of time $t$ \textemdash one day at a time: one day a unit time. Time $t$ the horizontal axis, the corresponding price $p$ the vertical one. We take the signal $(t,p)$ and project these discrete values onto the vertical axis $p$. This set of $1000$ points is a cantordust contained in a vertical segment $S$. We cover vertical $S$ with adjacent interval-boxes of equal size $l$. The number $L$ of boxes should not exceed, ideally, the square root of the total $T=1000$ points in the signal. Each point in vertical $S$ is a certain market price: the higher the point in $S$ \textemdash i.e. the higher the box $l_i$ that contains it\textemdash \space the higher the price $p$. The weight $w_i$ of box $l_i$ is the number of days in which market prices $p$ fell between the vertical bounds or extremes of $l_i$. Thus, each such box represents an increment $\Delta p$ in prices $p$. Let $p^-$ and $p^+$ be the extremes of segment $S$: the lower and upper prices. For NQ1000 as well as for the DJIA's, we always notice a rather small set $s$ of very heavy boxes quite near $p^-$. These $s$ heavy boxes represent many days in which market prices stood firm, with constant values, i.e. prices rather stable, not very expensive. The short lower subsegment in $S$ between $s$ and $p^-$ represents prices going down \textemdash hence its smallness: we will do without it, as generally, prices tend to go up. Hence, we will work  with heavy boxes as the lowest $l_i$ boxes, and study the nature of price increase. The remaining vertical upper segment in $S$, from heaviest boxes at the bottom to the lightest ones on top, will be normalized to be the unit segment $[0,1]=I$, the cantordust thus reduced will be \emph{the} fractal cantordust $K$. Since the signal has a total $T$ of 1000 points, then $L$ \textemdash the number of $l_i$ boxes\textemdash \space should not exceed $\sqrt{1000} \approx 32$. The sum of all weights in all boxes is 1000. The situation at the top, near $p^+$, presents a problem: there are boxes with just 1 point, on top of which we find a box with 0 points, then, on top, another box with 1 point\dots These isolated \textemdash and extremely light boxes\textemdash \space correspond to prices quite high\dots and most unstable, for they came to pass during 1 day (the weight $w_i$ of that $l_i$ box) only, a situation made worse by the upper \emph{empty} box, followed by an even \emph{higher} box again with 1 point \textemdash even higher market price, during one day, with an increase of $2\Delta p$: skipping the empty box\dots These abrupt jumps in prices not only reek of market crash, they pose a geometrical problem, for isolated and far away points do not belong to a fractal \dots a problem we will deal with below. The boxes at the bottom of vertical $I$, being the heaviest, correspond to $\alpha_{min}$, the highest ones, being the lightest, to $\alpha_{max}$. 
\newline\indent
We have a number of things superimposed, illustrated in diagram D (Fig. 1) vertical cantordust $K$ in $I$, the sequence of price increases $\Delta p$ (one such increase for each box $l_i$ with weight $w_i$), vertical axis $p$ with stable price for lower box and higher price for upper one, and the corresponding vertical interval $[\alpha_{min},\alpha_{max}]$, with $\alpha(f_{max})$ somewhere in the middle.
\newline\indent
We will concentrate our study of market crash situation in two separate parts: the analysis of what happens between $\alpha_{min}$ and $\alpha(f_{max})$, and  the study of the descending branch \textemdash $\alpha(f_{max})$ and $\alpha_{max}$\textemdash \space of the spectrum.
\newline\indent
The number of adjacent intervals $\Delta \alpha$, of equal size, covering vertical interval $[\alpha_{min},\alpha_{max}]$ should \textemdash ideally\textemdash \space not exceed the square root of the number $L$ of boxes $l_i$ covering $I$. That leaves us with six such intervals $\Delta\alpha$, four of which cover the interval $[\alpha_{min},\alpha(f_{max})]$.
\section{Between $\alpha_{min}$ and $\alpha(f_{max})$}
The value $\alpha(f_{max})$ is calculated in two ways that agree: it is the value of $\alpha$ for which $q=f'(\alpha)=0$ in the lagrangian coordinates. On the other hand, going to the spectrum-by-definition, it is in the fourth interval $\Delta\alpha$ that all values of $\alpha$ in that interval cram together around the value for which $q=f'(\alpha)=0$, as observed in a histogram that contains \emph{all} $\alpha$'s\textemdash one for each box $l_i$. $N_\alpha$ for that particular $\Delta\alpha$ is no smaller than for any other $\Delta\alpha$, also: its $\alpha$ values are the only ones quite near one another, crowded around the value of $\alpha$ for which $q=0$.
\newline\indent
In the symmetric case above we had a decreasing sequence of weights $w_i= P^k,P^{k-1}, \dots P^{k-r}$ \dots with $P>1$, each weight corresponded to a certain $\Delta\alpha$, each $\Delta\alpha$ was reduced to a point in that ideal case, with $N_\alpha= {k \choose r}$ boxes $l_i$. 
\newline\indent
Here we have the lowest $\Delta\alpha$, corresponding to $\alpha_{min}$, endowed with $N_\alpha=4$ such boxes with weights $154,142,84$ and $72$ days, respectively, with an average of $114$ days, or $3.5$ months. In the same way in which we identify $\alpha$'s in a certain $\Delta\alpha$ \textemdash $N_\alpha$ the number of such $\alpha$'s\textemdash \space we will average weights $w$'s of boxes $l_i$ in that $\Delta\alpha$. We recall that each $l_i$ box \textemdash pertaining to an interval $\Delta\alpha$\textemdash \space represents a price increase $\Delta p$ inside pre-determined bounds $[p,p']$ given precisely by the extremes of interval $\Delta\alpha$. 
\newline\indent
The second $\Delta\alpha$, immediately above the lower $\Delta\alpha_{min}$, has $N_\alpha=7$ boxes $l_i$ with weights $64,61,50,42,42,42$ and $31$ days, respectively, with an average of $47$ days or approx. 1.5 months. The third $\Delta\alpha$ above has an average of 20 days or three weeks, and the fourth one, reaching $\alpha(f_{max})$, an average of 10 days. So we have a sequence of $\approx 3.5$ months, slightly more than $1.5$ months or $6$ weeks, $\approx 3$ weeks or 20 days, 10 days: it does look like $P^k,P^{k-1},\dots,P>1$ with $k=6$ and $P$ slightly larger than $2$ \textemdash indeed, the average of these "$P$'s" is $\approx 2.2$; each $P=\frac{P^{k-r}}{P^{k-(r+1)}}$ is taken as the ratio of two consecutive average number of days above. (In fact, the two remaining $\Delta\alpha$'s between $\alpha(f_{max})$ and $\alpha_{max}$ do continue with this division by $\approx 2$: the average of 5th interval $\Delta\alpha$ is slightly less than $5$ days, and the last one corresponds to the 1 or 2 days per box.) 
\newline\indent
We can now proceed in \emph{three different ways} that indicate an early warning of the market crash situation. We call it "early" because $\alpha(f_{max})$ is situated far below $\alpha_{max}$, and $\alpha_{max}$ is at the very top of the market price list: $p^+$. So $\alpha(f_{max})$ is midway between $p^-$ and $p^+$, far below $p^+$. The logic is that before prices reached $p^+$ values, on their way up, they passed through these critical values corresponding to $\alpha(f_{max})$: hence, what happens at this value is an \emph{early} warning of future market collapse.
\subsection{The instability index}
In Sec. 2.1 we defined the instability index $Q_i(r) = \frac{N_\alpha}{P^{k-r}}, r =0,1,2,\dots$ We recall that $P>1$, that $k$ was fixed, that $\alpha=Ar+B$, so we can write $Q_i(\alpha)$ or $Q_i(r)$; that for the binomial case $Q_i(r)$ is steadily increasing with $r$ (or $\alpha$), and that we are between $\alpha_{min}(r=0)$ and $\alpha(f_{max})$, which means $r=\frac{k}{2}$. In this last ideal case, $Q_i(r)$ grows with no change of curvature between $r=0$ and $r=\frac{k}{2}$ (Fig. 2, out of scale). In the case of our NASDAQ signal, the weights $P^k,P^{k-1},\dots$ are $114,47,20$ and $10$ days (Sec. 4, with a value of $P\approx 2.2)$ until we reach $\alpha(f_{max})$; the four $N_\alpha$ being $4,7,5$, and $7$ boxes in the four intervals $\Delta\alpha$ (Sec. 3) between $\alpha_{min}$ and $\alpha(f_{max})$. The graph $Q_i(r)$, in this case, plotted against four integer values of $r$, shown in Fig. 3 (out of scale), has an inflection point, between $\alpha_{min}$ and $\alpha(f_{max})$, near the latter.
\newline
\underline{Note:} the values $N^1_\alpha=4,N^2_\alpha=7,N^3_\alpha=5, N^4_\alpha=7$, do not imply that the second and fourth $\Delta\alpha$ have maxima, for in the fourth $\Delta\alpha$ the $\alpha$'s are concentrated around $f'(\alpha)=0$, whereas in the second $\Delta\alpha$ they are dispersed. 
\newline\indent
Let us interpet $Q_i$. The first $Q^1_i=\frac{4}{114}=0.035$ is much smaller than the second one: $\frac{7}{47}=0.148$: the instability increased. In the first instance, prices increased every 3.5 months, a rather stable situation, and they did so for 4 consecutive $\Delta p$ price increases. Next, the prices began to increase $\Delta p$ each month and a half, a situation that repeated itself seven times, that is why $Q_i$ increased. The inflection point above is due to an anomaly in the evolution of the market: 5 consecutive price increases $\Delta p$ each 3 weeks are immediately followed  by price increases each 10 days only! \dots and this abnormal jump takes place 7 consecutive times\dots The logic underneath is the following: we had a price increase each month and a half, 7 times. Next, we find a $\Delta p$ increase each 3 weeks, which is a much worse situation. But, if this situation is repeated a rather-large-than-7 times \emph{instead of 5 times}, there is room for the system to adapt, there is valuable time to adjust to the new situation \emph{before} time intervals between consecutive $\Delta p$ shorten again. This numerical anomaly in the values of $N_\alpha$ causes the inflection point in the graph of $Q_i(\alpha)$.
\newline\indent
So, real stability of the system does depend on $Q_i$ remaining low, but if it grows, the occurrence of a crash depends on the geometry of its graph (inflection point), as described above. 
\subsection{The second way to study crash early warning: the lagrangian spectrum}
We continue with NASDAQ, we take boxes $l_i$ with weight $w_i$, divide each $w_i$ by $T=1000$ and obtain probabilities $p_{r_i}$. Then we can apply the thermodynamical algorithm and obtain $f(\alpha)$, as shown in Fig 4. Two observations: a) the curve $(\alpha,f(\alpha))$ is \emph{not} symmetric \textemdash as is the binomial one; b) there is a noticeable gap between $f'(\alpha)=1$ and $f'(\alpha)=0$ \textemdash entirely in the interval $[\alpha_{min},\alpha(f_{max})]$. 
\newline\indent
Since $f'(\alpha)=0$ means $\alpha=\alpha(f_{max})$, we observe that the inflection point of $Q_i$, very near $\alpha(f_{max})$, is located precisely in that gap.
\begin{figure}
\begin{center}
\includegraphics[width=100mm]{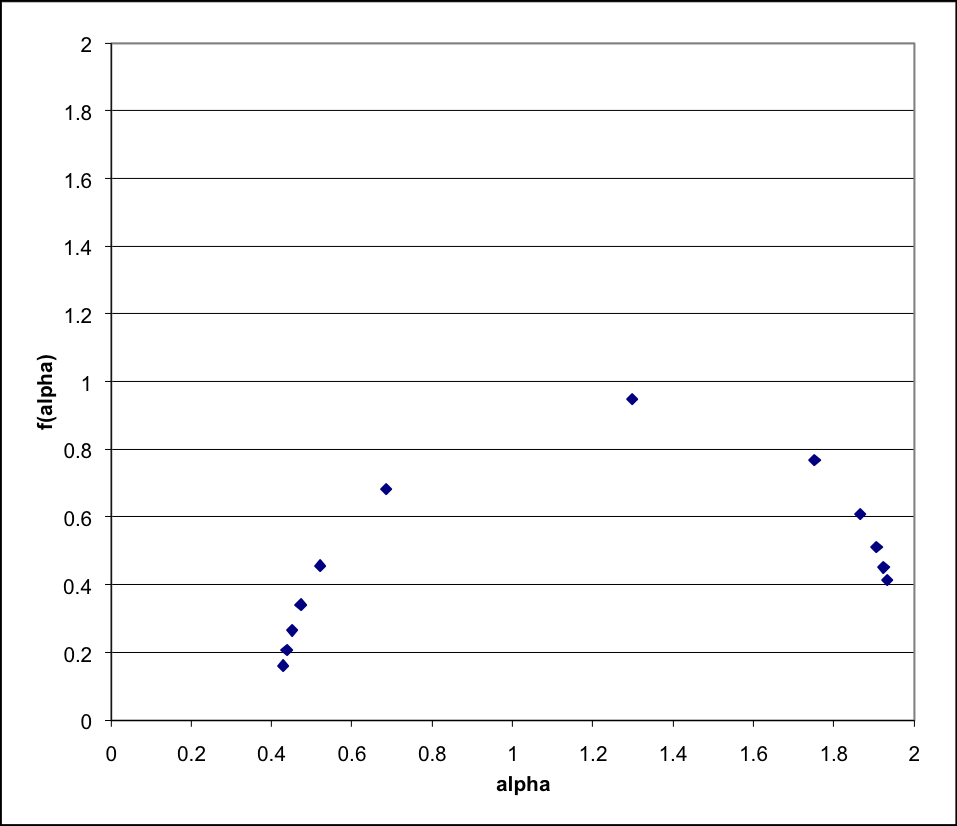}
\caption{Thermodynamical formalism, NQ1000}
\end{center}
\end{figure}
\begin{figure}
\begin{center}
\includegraphics[width=100mm]{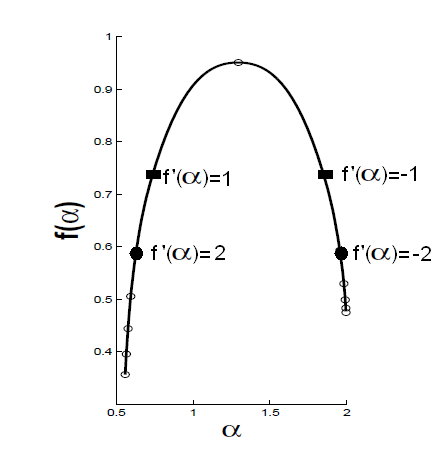}
\caption{Symmetric thermodynamical formalism, non-crash situation}
\end{center}
\end{figure}
\begin{figure}
\begin{center}
\includegraphics[width=100mm]{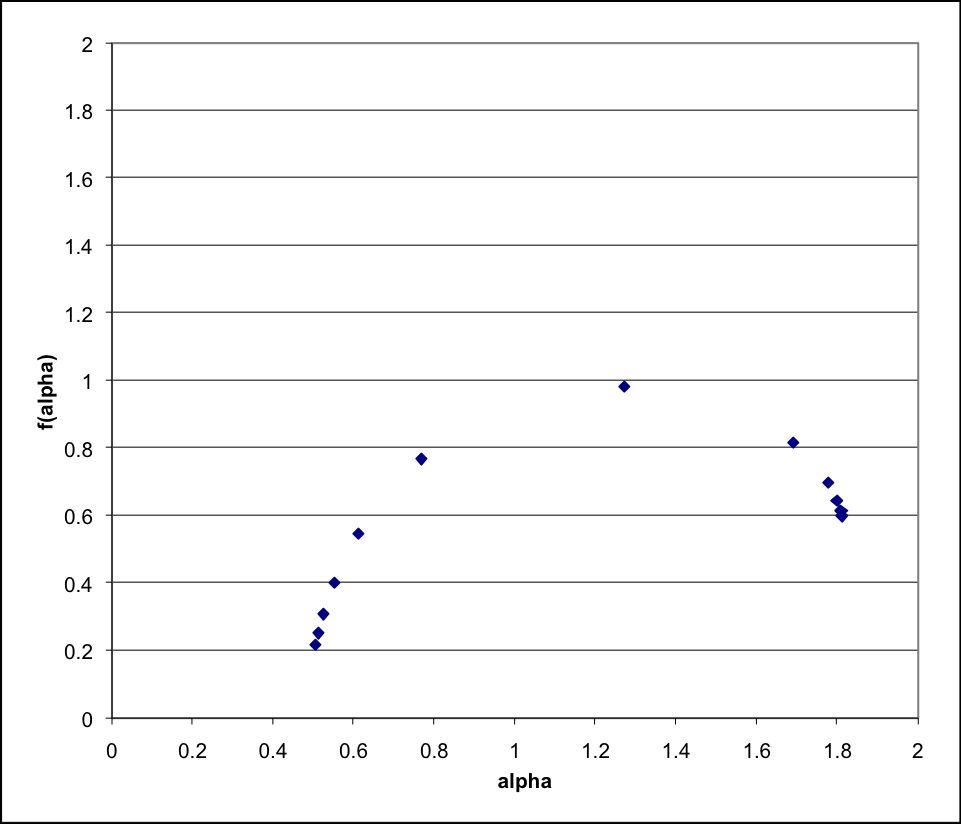}
\caption{Thermodynamical formalism, DJIA1600}
\end{center}
\end{figure}
If we consider a time span where there is no important crash, e.g. the daily fluctuations of Dow Jones from 1st Jan. 1990 to 1st Jan. 1994, the corresponding curve $(\alpha,f(\alpha))$ \emph{is} symmetric, as shown in Fig. 5. It should be noticed that symmetric spectra are quite rare. This case is, as far as we are aware, unique in empirical context \textemdash the binomial case is a purely mathematical one. Moreover, if we take the data from the  DJIA1600 sample, still the corresponding $(\alpha,f(\alpha))$ curve mantains both the gap \emph{and} the asymmetry (Fig. 6): enlarging the prior-to-crash sample does not "dilute" these crash features. 
\newline\indent
Further, if we do the opposite, that is, take smaller and smaller samples prior to crash: DJIA900, 800, 700, 600, always maintaining $L=\sqrt{T}$ ($L$ = number of $l_i$ boxes, $T$ = size of the sample, all as above), we obtain \emph{the same} $f(\alpha)$ as in Fig. 4 \textemdash with minute variations (Fig. 7): the process is \emph{strictly} self similar (i.e. independent of size sample, or of scale) around market crash point. 
\begin{figure}
\begin{center}
\includegraphics[width=100mm]{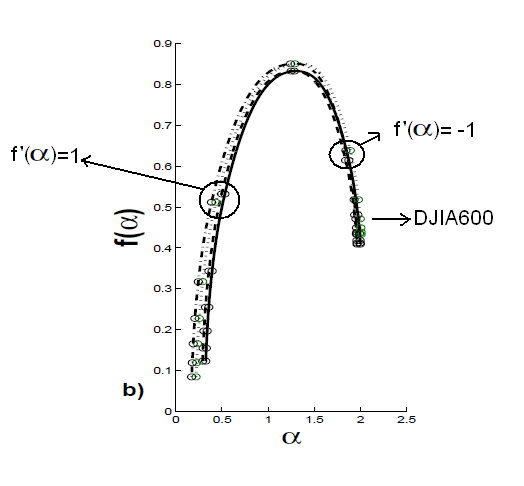}
\caption{Thermodynamical spectra for DJIA900, 800, 700, 600}
\end{center}
\end{figure}
\underline{Note}: In order to plot the thermodynamical lagrangian spectra curve $(\alpha,f(\alpha))$ in each case throughout this section, we have used only integer values for $q=f'(\alpha)$, since they are enough to determine the key properties of $f(\alpha)$: the graph dots for $q=0, +\frac{1}{2}, -\frac{1}{2}, +1, -1, +\frac{3}{2}, -\frac{3}{2},\dots$, would be deceiving, since the corresponding graph points at the extremes of the curve would simply pile up on top of each other, hence the upper gaps would fill up artificially, without adding any real meaning to the geometry of the lagrangian process.
\subsection{A third way to study the early warning gap: the $f^*(\alpha)$  spectrum}
We continue with NASDAQ data. Recall that $T=1000$, and that we have between 30 and 40 $\alpha$'s, from boxes $l_i$ grouped in 6 intervals $\Delta\alpha$, the first four span $[\alpha_{min},\alpha(f_{max})]$, the $\alpha$-interval of our study. We also recall that $f^*(\alpha)$ is the spectrum-by-definition and $f(\alpha)$ the lagrangian one. The first $\Delta\alpha$ interval, the one corresponding to $\alpha_{min}$, has 4 $\alpha$'s, more or less distributed evenly in the interval. The second adjacent $\Delta\alpha$ (seven $\alpha$'s) has its $\alpha$'s not so dispersed, some are near the $\alpha$ for which $q=f'(\alpha)=1$. The third $\Delta\alpha$ has very few and dispersed $\alpha$'s. So $f^*(\alpha)$ and $f(\alpha)$ share the same $\alpha_{min}$, then $f^*(\alpha)$ grows to a local maximum at an $\alpha$ for which $f'(\alpha)=1$,\dots then it goes \emph{down}, and it does it in the gap between $f'(\alpha)=1$ and $f'(\alpha)=0$. The fourth interval $\Delta\alpha$ includes the particular $\alpha$ for which $f'(\alpha)=0$. All its $\alpha$'s are clustered around this particular $\alpha$, which implies that a spectrum (another one) $f^*$ should have an \emph{absolute maximum} at that value of $\alpha$, just as $f(\alpha)$ does \textemdash in the 5th and 6th remaining $\Delta\alpha$ intervals the number of $\alpha$'s steadily decreases, and so does this last $f^*(\alpha)$. 
\newline\indent
The picture corresponds to \emph{two} $f^*(\alpha)$: $f^*_1(\alpha)$ starts in $\alpha_{min}$, reaches a maximum $\alpha_1$ for $q=f'(\alpha)=1$, and descends in the gap between $q=1$ and $q=0$. The other, $f^*_2(\alpha)$, goes from its own maximum at $\alpha_0$ (for which $q=f'(\alpha)=0$) and decreases as $\alpha$ nears $\alpha_{max}$ (shared by $f^*_2(\alpha)$ and $f(\alpha)$). By considering a larger sample prior to market crash, DJIA1600, we have one more $\Delta\alpha$, which clarifies the situation: $f^*_2(\alpha)$ starts with a \emph{low} value in the gap between $\alpha_1$ and $\alpha_0$ (i.e. $q=1$ and $q=0$), then grows to an absolute maximum at $\alpha_0$, where $f'(\alpha)=q=0$, and then decreases as $\alpha$ nears $\alpha_{max}$: clearly, two different $f^*_{1,2}(\alpha)$ (with negative second derivative), which seem to collide in "the gap"; Fig. 8 shows a schematic diagram representing this situation. The two  $f^*_{1,2}(\alpha)$ multifractal spectra \textemdash bi-multifractality\textemdash  \space are two curves for which $f(\alpha)$ acts as an envelope, as a smooth upper "roof" of both. According to Radon and Stoop (1996), this situation \textemdash bi-multifractality with an upper envelope\textemdash \space occurs when \emph{two} different fractal measures are processed with \emph{one and the same} multifractal algorithm. 
\begin{figure}
\begin{center}
\includegraphics[width=100mm]{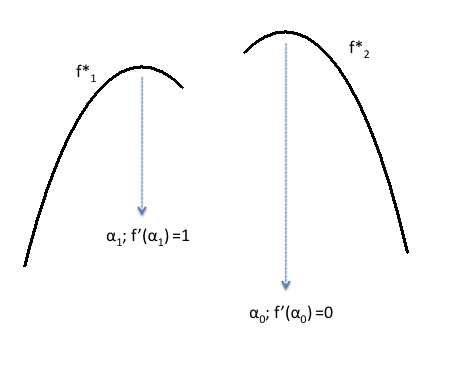}
\caption{Diagrammatic spectra $f_{1}^*$ and $f_{2}^*$}
\end{center}
\end{figure}
\section{The situation between $f'(\alpha)=0$ and $\alpha_{max}$}
As $\alpha$ approaches $\alpha_{max}$, the market prices increase more and more, to the point in which they can be sustained during one day only. This situation, as we noticed before, is reflected in boxes with 1 or 0 points (days) near the $p^+$ top. We have several such boxes at the top prices, separated from each other by zero point empty boxes. This situation, which we call "dispersion", depicts a number of isolated points far away from the main body of the fractal cantordust, and, from the point of view of fractal geometry, they should be discarded \textemdash no isolated point contributes dimensionally: that is why we used several times the expression "between 30 and 40 boxes", for quite some such boxes have to be discarded when doing a spectrum-by-definition $f^*(\alpha)$. Instead, the lagrangian $f(\alpha)$ process does not seem to care \emph{where} the boxes are: isolated and far away or in the very midst of the cantordust, but \emph{how many} they are, for each value of $\alpha$, regardless of their location. Of these two spectra, we choose $f(\alpha)$ in order to interpret and quantify the dispersion, since it takes into consideration \emph{all} the dispersed boxes near $\alpha_{max}$. Near crash, prices jump so high, and so high, that we have many dispersed boxes: $N_{\alpha_{max}}$ is large, and so is $f(\alpha_{max})$, as we observe for NASDAQ crash in Fig. 4, where the large dispersion in $\alpha_{max}$ substantially contributes to the non-symmetry of the curve. 
\newline\indent
Carefully notice that, when we enlarge the prior-to-crash sample, e.g. DJIA 1600, Fig. 6, we slightly "dilute" the crash effect on the descending right branch of the corresponding $f(\alpha)$, for $f(\alpha_{max})$ is much smaller \textemdash and the dots closing the right branch are much more separated than in the NASDAQ case, though they are drawn with the same parametric $q$ values as for NASDAQ. Also, Fig. 7 shows that as we take the sample nearer and nearer crashpoint, DJIA900, 800, 700, 600,\dots the higher the value of $f(\alpha_{max})$: the highest one corresponds to DJIA600 (the numerically smallest sample which still yields a cantordust). The economic interpretation: those extraordinarily high jumps in market prices \textemdash the dispersion\textemdash \space increase as the system destabilizes, as we get dangerously near its collapse.
\newline\indent
\underline{Note}: An observation on the inflection point in $Q_i$ (Fig. 3): $Q_i$ grows between $\alpha_{min}$ and $\alpha(f_{max})$, since the numerator $N_{\alpha}$ grows, and the denominator is a decreasing power. When $f^*_1$ grows (up to $q=1$) then $Q_i$ grows, ditto when  $f^*_2$ grows \dots but, in the middle of these two growing spectra, exactly between $q=1$ and $q=0$, $Q_i$ does not grow \dots but sits down, hence the inflection.
\section{Conclusions}
A prior-to-crash  market signal is studied as a cantordust, its fractal properties, responsible for the crash situation, are studied with the tools of multifractal analysis. Two different multifractal spectra $(\alpha,f(\alpha))$, and an instability quotient $Q_i$ characterize and quantify an "early warning" of market collapse, with market prices (not yet very high) corresponding to values of $\alpha$ between $\alpha_{min}$ and $\alpha(f_{max})$; the study of the highest market prices, corresponding to values of $\alpha$ near $\alpha_{max}$, is done by analysing the fractal dispersion of the cantordust prior-to-crash market signal. 

\section{References}
Radon, G. and Stoop, R. (1996). "Superpositons of multifractals: generator of phase transition in the generalized thermodynamic formalism". J. Stat. Phys., Vol. 82, Nos. 3/4, 163-180.
\newline\newline
Rotundo G. (2006). "Logistic Function in Large Financial Crashes" in Logistic Map and the Route to Chaos. Springer NY.

\end{document}